# A method for coupling dynamical and collisional evolution of dust in circumstellar disks: the effect of a dead zone.

Sébastien CHARNOZ[1,2,3]

Esther TAILLIFET[1]

(1) Laboratoire AIM, Université Paris Diderot / CEA /CNRS
(2) Institut Universitaire de France
(3) Corresponding author: charnoz@cea.fr

**Abstract**

**Dust is a major component of protoplanetary and debris disks as it is the main observable signature of planetary formation. However, since dust dynamics is size-dependent (because of gas-drag or radiation pressure) any attempt to understand the full dynamical evolution of circumstellar dusty-disks that neglect the coupling of collisional evolution with dynamical evolution is thwarted because of the feedback between these two processes. Here, a new hybrid lagrangian/eulerian code is presented that overcomes some of these difficulties. The particles representing "dust-clouds" are tracked individually in a lagrangian way. This system is then mapped on an eulerian spatial grid, inside the cells of which the local collisional evolutions are computed. Finally, the system is remapped back in a collection of discrete lagrangian particles keeping constant their number. An application example on dust growth in a turbulent protoplanetary disk at 1 AU is presented. First the growth of dust is considered in the absence of a dead-zone and the vertical distribution of dust is self-consistently computed. It is found that the mass is rapidly dominated by particles about a fraction of millimeter in size. Then the same case with an embedded dead-zone is investigated and It is found that coagulation is much more efficient and produces, in a short time scale, 1cm-10cm dust pebbles that dominate the mass. These pebbles may then be accumulated into embryos sized objects inside large-scale turbulent structures as shown recently (see e.g. Johansen et al., 2007).**

# Introduction

Circumstellar disks, be they young protoplanetary disks or evolved debris disks, are complex systems that are detected in the infrared through the emission of the smallest dust particles they contain. Learning about the origin and the fate of this dust is of crucial importance since the observable dust is commonly used as a proxy to quantify the unseen, but suspected, population of macroscopic bodies; either planetesimals, planetary embryos or planets. To interpret observations and link the visible dust population to the unseen population of planetary bodies, information on the particle size-distribution is necessary. In the last decades, numerous efforts have been made to compute self-consistently the size distribution of dust in protoplanetary disks (i.e. gas rich) or in debris disks (i.e. gas poor) (see e.g. Brauer et al., 2008 or Thébault 2011 and references therein). However, the main obstacle that prevented almost any attempt to do such a computation is the difficult coupling of the dust *dynamical* evolution with the dust *collisional* evolution. In most of works, either one aspect or the other is studied and only a handful of papers (see e.g. Charnoz and Morbidelli 2004, Brauer et al., 2008; Stark and Kushner 2008, Thébault 2011) have attempted to tackle this difficult problem, but succeeded under some restrictive assumptions only. So there is a need for a more general method with the least possible assumptions, which is the subject of the present paper.

**1.1 Lagrangian vs. Eulerian**

The wish of any scientist computing collisional evolutions of dusty protoplanetary or debris disks would be to have a dynamical code (like an N-body code) to compute the local dynamics and the local encounter velocities, then use these encounter velocities inside a Particle In A Box (PIAB) code to evolve the dust size distributions. However simple, and not new (see e.g. Charnoz and Morbidelli 2004, 2007 or Stark and Kuchner 2008 etc.) this method is difficult to put in practice because of the intrinsically different natures of N-body and PIAB codes. Indeed, dynamical N-body codes are often *lagrangian* in their essence so that a dust grain is represented by an individual particle whose motion is tracked, so it is perfectly suited to compute the velocities. Conversely PIAB codes use local ensemble averages, so they are *eulerian* in their nature, and local quantities (mass, number of particles etc…) are transported from one cell to another in the phase space. In consequence, they are perfect to compute the average evolution of large ensemble of particles, like the size-distributions. Because of a lack of computer power (we need to track a system with 7 dimensions, 3 for positions, 3 for velocities, 1 for particle size) PIAB methods are not used to track the system's dynamics or are, at best, very inaccurate.

**1.2 Different attempts for different physics**

Among the attempts to couple dynamical and collisional evolution of interplanetary bodies, the followings different methods may be noted. Charnoz and Morbidelli (2004, 2007) tracked the collisional and dynamical evolution of the Kuiper Belt and the Oort cloud including giant planets' gravitational perturbations by using super-particles and a N-body code. However the method they used is not directly applicable to dust as it relies on a size-independent dynamics (so while they compute the evolution of size distribution, they neglect the effect of dissipative rebounds, but justifying it at posteriori), which is only valid for macroscopic bodies (> 100 µm). For gaseous and dusty protoplanetary disks, Brauer et al. (2008) and Birnstiel et al. (2010) coupled the dynamical and collisional evolution of dust in the disk's midplane using analytical prescriptions for dust dynamics. They assumed instantaneous vertical steady state and instantaneous coupling to the gas. This model should give a good representation of the dust physics in the disks' midplane. However it may be largely inaccurate in the disk's upper layers that are, unfortunately, the only part accessible to IR observation. For debris disk, Grigorieva et al. (2007) proposed a method close to a N-body code that creates new particles at each time-step, because of fragmentation processes. The drawback of this method is that the number of particles increases exponentially with time, restricting that study to short timescales, about an orbital period only. More recently, for debris disks, Stark & Kuchner (2008) presented a method by which they produced maps of the steady state dynamics for all dust sizes particles. They weighed the different maps according to the evolution of the size distribution. In a way, the velocity field is pre-computed for each size bin and used as an input for a collisional code. This method is particularly adapted for systems close to steady state, but cannot be used for systems following a time-dependent dynamics. In the same spirit, Thebault (2011) proposed a modified version of this approach for binary systems. Those two papers present maybe the most advanced technics for dust particles: they are efficient for steady state systems, but, up to now, it was not possible to study a time-dependent dynamical system with a complex dynamic. Note also that some works include the dynamical effects of collisions (dissipative rebounds among particles) but neglect fragmentation and coagulation. They use the method of inflated particles, as in Charnoz et al., (2001) did in the context of a planetesimals disk , or recently in Youdin et al. (2012) in the context of a mixture of dust with gas (to study the streaming instability). Ultimately rebound, coagulation and fragmentation should be coupled. Nevertheless we limit ourselves, in the present paper, to coagulation and fragmentation: Because of intense gas drag in a protoplanetary disk, the dynamical effect of collisions is rapidly erased (for small particles, see below). However rebounds should be included in the future.

The method presented here shares some similarities (but is still largely different) with the algorithms presented in Ormel and Spaans (2008) or Zsom and Dullemond (2008). Like them statistical averages are used to compute pairwise collisions between small groups of representative particles. However, our method uses a much more refined dynamical model and the "remapping" algorithm allows keeping an optimal resolution in either both particle sizes and spatial resolution.

### 1.3 Structure of the paper

The present paper describes a method to couple a dynamical and a PIAB code in order to compute the coupled dynamical and erosional evolution of circumstellar disks with a time-dependent dynamic in 3D. The main ingredients that enter in our code and make the computation traceable are the following:

- The use of super-particles called "tracers", that each represents a cloud of many dust grains of a single size rather than a single body.

- The use of a "collisional grid" on top of the system, in the cells of which the local collisional evolutions are computed using a PIAB method.

- A "remapping" algorithm to reorganize tracers at each time-step in order to present the system in an optimal way with a constant number of tracers.

The method presented here is flexible and could be, in principle, applied to any system provided there are enough particles and cells.

The method used in the code is presented in details in section 2. In section 3 examples of applications are presented. In section 3.1 a time-step is decomposed in its different phases. In section 3.2 a comparison with the work of Brauer et al. (2008) is presented. As an example, section 3.3 compares the dust growth in disks with and without a dead-zone.

As opposed to what is often assumed for simplicity, it is shown that the vertical distribution of particles is not at dynamical equilibrium, because of the very short collisional timescale. Consider the case of a protoplanetary disk containing a dead-zone it is shown that coagulation processes rapidly produce big particles, up to 10cm in size, close to the disk's midplane. The case of a debris disk, that has unique specificities (like particles escaping from the system) will be the subject of a companion paper as will the case of a full disk extending from 1 to 100 AU.

This code is based on a recently published code LIDT3D (for Lagrangian Implicit Dust Transport in 3D, Charnoz et al., 2011) designed to track individual dust particles in a turbulent disk.

## 2. Description of the LIDT3D code

In this section, the different steps of the code are presented on a theoretical point of view. A step-by-step illustration is given in section 3.1.

### 2.1 General overview and principles

The system is made of a collection of super-particles called "tracers" that evolve dynamically like in a classical N-body code. Each tracer represents a cloud of dust grains of a single size. These tracers constitute a lagrangian description of the system. At each time-step the system is shifted from the lagrangian description (made of tracers, adapted to compute the dynamical evolution) to a eulerian description (using a spatial grid, adapted to compute the evolution of size distributions). We call this a "re-mapping", because the system is remapped back-and-forth from one description to another at each time-step. This is illustrated in Figure 1. First, the tracers' motions are computed (Figure 1.a). Then the tracers are mapped onto a grid (Figure 1.b). The collisional evolution is computed using a PIAB collisional model inside each grid's cell. The system is then remapped-back to the lagrangian description (using tracers only, Figure 1.c). During this phase, tracers are spatially re-organized (Figure 1.d) and may be moved from one cell to another in order to map, in the optimal way, the new spatial distribution of dust-grains with their different sizes. To ensure mass conservation, when a tracer is moved, it transfers to its neighbor tracers tracing the same grain size its dust-mass content. This way the local mass is conserved and not altered by the reorganization. The empty tracer is then attributed to a new position, with a new velocity and a new dust content corresponding to the local state of the system at its new location.

These steps are detailed below.

### 2.1 Definition of a tracer

Tracers are the elemental bricks of LIDT3D. Each tracer is a lagrangian particle associated to a position and a velocity that represents a collection of dust grains of a same size. The system is represented by N tracers (N≥10,000 typically). The tracer numbered *p* is given a position vector **$R_p$**, velocity vector **$V_p$**, radius $r_p$ and dynamical mass $m_p$. All dust grains in the dust cloud represented by tracer *p* have the same mass $m_p$ and radius $r_p$. The total mass of dust contained in the dust cloud is $M_p$ and the total number of dust grains is $N_p$, so that the following relation is always true for any tracer *p* : $M_p = N_p \cdot m_p$ and $m_p = 4/3\pi \rho \times r_p^3$ where ρ is the mass density of the dust's material.

A finite number $N_s$ of different dust size-bins are considered, from an arbitrary minimal radius ($r_{min}$) to an arbitrary maximal radius ($r_{max}$) with logarithmic increments. Since a tracer represents a collection of same size dust grains, a tracer p is associated with a size bin j (with $1<j<N_s$) that depends on $r_p$ value. A size bin corresponds to a certain range of dust sizes. Note that the size bin to which a tracer belongs may change during a simulation. At the end of the remapping phase (see section 2.5), a tracer can change position. Its size bin may change, as well as its position and velocity, depending on the local size distribution of dust at the new location.

A time-step in LIDT3D is divided in 4 steps, described in details below.

**2.2 Step 1 : Dynamical evolution of tracers from t to t+dt (Figure 1.a)**

Each tracer evolves like a particle with mass $m_p$ and radius $r_p$ in the force field of the system (see Figure 1.a). In a protoplanetary disk, a tracer moves as a result of the central star's gravity (with mass $M_*$) and the drag induced by the surrounding gas (with mass density $\rho_g$, sound velocity $C_s$ and local velocity vector $\vec{V}_g$). When the gas drag is in the Epstein regime the braking time is $\tau = r_p \rho / C_s \rho_g$. Thus, the equations governing a tracer's motion are:

$$\begin{cases} \dfrac{d\vec{R}_p}{dt} = \vec{V}_p \\ \dfrac{d\vec{V}_p}{dt} = -\dfrac{GM_*}{a_p^3}\vec{R}_p - \dfrac{\vec{V}_p - \vec{V}_g}{\tau} \end{cases} \quad [1]$$

In Eq. [1] $a_p$ is the distance of a tracer to the central star and $M_*$ is the star's mass. In the case of a debris disk, radiation pressure should be added to Eq. [1] while gas-drag may not be considered. The action of a perturber (like a planet) can be also added to Eq. [1] if necessary.

Any ODE solver can be used to solve Eq.[1]. However an implicit method is highly recommended here since Eq. [1] is a stiff equation because of the two very different timescales at play: the stopping timescale $\tau$ and the orbital timescale $T_k$. If an explicit solver (such as the popular explicit fourth-order Runge-Kutta scheme) were used, the integration time-step would be limited to a fraction of the gas-coupling timescale $\tau$. We would be constrained to leave out the smallest dust grains or to integrate their motion over short timescales.

In the present work a Bulirsch-Stoer scheme is used, with a semi-implicit solver (Bader & Deuflhard 1983) as described in Press et al. (1992, Chapter 16.6) that gives excellent results for all particle sizes in a protoplanetary disk (see Charnoz et al., 2011 for more details).

**2.3 Step 2: Remapping tracers onto a grid, computing local velocities (Figure 1.b)**

To compute the evolution of size distributions using a PIAB algorithm, a spatial grid is defined on top of the system. In each cell of this grid the local average encounter velocities will be computed as well as the evolution of the dust size distribution (see Figure 1.b). Because of the lack of computer power a two-dimensional grid is considered for now. However it is straightforward to extend our method to the 3D case (note that the tracers' motion is always integrated in 3D). The grid's geometry should depend on the problem. For a protoplanetary disk in which it is interesting to compute the evolution of dust size distribution as a function of R (the distance to the star) and Z (the distance to the midplane) an (R,Z) grid is adapted (assuming azimuthal symmetry). For a debris-disk in which azimuthal structures must be resolved, a polar grid (R,θ) may be a better choice.

Whatever the choice of grid, in each cell and for every time step, it is assumed that the dust clouds (represented by tracers) collide with each other. The encounter velocities need to be calculated for the different pairs of tracers representing clouds of different dust sizes present is the cell. Those velocities will then be considered as constant in this cell. In other words, variations of encounter velocities' inside a single cell are neglected during a time-step dt.

Let us consider a cell of identification index k. In order to calculate the encounter velocities we first need to identify the tracers present in the cell k at the end of the step 1. Let *p* the indexes referring to the tracers inside the cell *k* that belong to the bin size *j*. The dust mass distribution inside cell k is obtained by summing the mass of all tracers inside cell k for all size bins j. Let $M_{j,k}$ (with $1<j<N_s$) the array containing the dust mass in each size bin j inside cell k and $N_{j,k}$ (with $1<j<N_s$) the array containing the number of dust grains in each size bin j inside cell k, so that :

$$M_{j,k} = \sum_p M_p \qquad [2]$$

$$N_{j,k} = \sum_p N_p \qquad [3]$$

The average velocity vector for the size bin j in the cell k is called $<\mathbf{V}_j>_k$ (note that $<\mathbf{V}_j>_k$ is a 3D vector). Because of the finite radial size of the cells, $<\mathbf{V}_j>_k$ cannot be simply obtained by computing the local average of all tracers' velocity vectors belonging to cell k and bin size j. Indeed keplerian shear (i.e. the fact that orbital velocity decreases like $a^{-1/2}$, *a* being the tracer's semi-major axis) induces a systematic velocity difference for tracers with slightly different semi-major axes inside the same cell. Furthermore, particles on a same trajectory, but at different longitudes inside a same cell, may seem to have a non-zero encounter velocity because of the curvature of their orbit - although, if they were in conjunction, their true encounter velocity would be close to 0. To avoid these artifacts due the finite size of the cell the mean velocity difference are computed with respect to a well-known velocity field at the position of the particle: the local gas velocity or the local circular velocity. It has the advantage of substantially correcting the artifacts induced by both the keplerian shear and the velocity vector's orientation. It gives a reasonable approximation of the encounter velocities. The average velocity of tracers in size bin j belonging to cell k is computed as follows:

$$<\vec{V}_j>_k \approx \vec{V}_{k,g} + \frac{1}{Nt_{k,j}} \sum_p (\vec{V}_{p,j,k} - \vec{V}_{p,g}) \qquad [4]$$

Where $Nt_{k,j}$ is the number of tracers belonging to size bin j and cell k, $\mathbf{V}_{p,j,k}$ is the velocity vector of a tracer with identification number p that belongs to size bin j and cell k, $\mathbf{V}_{p,g}$ is the velocity vector of the gas at the location of tracer p and $\mathbf{V}_{k,g}$ the gas velocity at the center of cell k. Similar technics were used in Lyra et al. (2009) or in Johansen et at. (2012) for computing relative velocities between dust grains in the context of the simulation of the streaming instability. The dispersion velocity $<\sigma v_j^2>_k$ is obtained according to:

$$<\vec{\sigma v}^2{}_j>_k \approx \frac{1}{Nt_{k,j}} \sum_p \left[ \vec{V}_{p,j,k} - \vec{V}_{p,g} + \vec{V}_{k,g} - <\vec{V}_j>_k \right]^2 \qquad [5]$$

The average encounter velocity, $\Delta V_{i,j,k}$, between two populations with size bins i and j within a same cell k may be approximated by doing a simple quadratic sum as follows:

$$\Delta V_{i,j,k}{}^2 \approx \sum_{m=1}^{3} \left[ <V_{i,m}>_k - <V_{j,m}>_k \right]^2 + <\sigma v^2{}_{i,m}> + <\sigma v^2{}_{j,m}> \qquad [6]$$

Where subscript *m* designates the coordinate axis (x,y,z). Eq. [6] assumes that the dispersion velocities of particles i and j are not correlated. It is found to be a good first approximation in the case of protoplanetary disks.

Strictly speaking Eq. [6] allows us to estimate the relative velocities induced by the forces explicitly considered in Eq .[1]. However it may be useful to consider additional sources of random motions that cannot be explicitly considered in the dynamical model, like -for example- the thermal motion of the dust in the gas (that may not be negligible for the smallest particles) or the turbulent motion induced by gas-turbulence. These contributions may be envisioned as sub-grid effects like it is done, for example, in Brauer et al., (2008) or in Birnstiel et al., (2010). In the case of a turbulent protoplanetary disk this gives:

$$\Delta V_{i,j,k}{}^2 \approx \sum_{m=1}^{3} \left[ <V_{i,m}>_k - <V_{j,m}>_k \right]^2 + <\sigma v^2{}_{i,m}> + <\sigma v^2{}_{j,m}> \\ + \Delta V^2{}_{i,j,THERM} + \Delta V^2{}_{i,j,TURB} \qquad [7]$$

Where $\Delta V^2{}_{i,j,THERM}$ is the standard deviation of the dispersion velocities induced by the thermal agitations between particles in size bins *i* and *j*: $\Delta V^2{}_{i,j,THERM} = 8kT(m_i+m_j)/(\pi m_i m_j)$ (Brauer et al., 2008) where *k* is the Boltzmann's constant, *T* is the gas temperature and $m_i$ and $m_j$ are the mass of dust grains in size bins i and j . Similarly $\Delta V^2{}_{i,j,TURB}$ is the standard deviation of the dispersion velocities induced by the turbulent motion of the gas between particles in size bins *i* and *j*. $\Delta V^2{}_{i,j,m,TURB}$ may be calculated analytically using the closed formalism of Ormel and Cuzzi (2007).

**2.4 Step 3: Computing the evolution of the size distributions in each cell with a Particle In a Box formalism (Figure 1.c)**

The size distribution of the dust at the end of the time-step, after the dust clouds collided in cell k is computed as follows. Since the encounter velocities (the array $\Delta V_{i,j,k}$) were computed in Step 2, It is now possible to calculate $Nc_{i,j,k}$, the number of collisions between two populations of particles i and j inside cell k occurring during the time-step dt, so that :

$$Nc_{i,j,k} = \frac{\Delta V_{i,j,k} \cdot N_{i,k} \cdot N_{j,k} \cdot \pi (r_i + r_j)^2}{\Gamma_{i,j,k}} dt \qquad [8]$$

Where $\Gamma_{i,j,k}$ is the volume occupied by dust populations in size bins i and j in cell k. $\Gamma_{i,j,k}$ can in most cases be considered to be the volume of cell k. Thus, $\Gamma_{i,j,k}$ is the product the cell's surface $S_k$ by the cell's vertical height $\delta z_k$. However, this assumption may not be justified close to disk's midplane where the low inclination tracers can represent a volume smaller than a cell. This is why it is worth comparing the cell's scale height to the particles' scale height. The scale height occupied by particles in size bins j and in cell k can be approximated by the ratio of the vertical velocity to the orbital frequency of particles ($\Omega_k$) $\approx 2<V_j>_{k,z}/\Omega_k$ where $<V_j>_{k,z}$ is the mean vertical velocity of the tracers in size bin j and cell k, computed using Eq.[4]. The expression for the effective volume of interactions is then:

$$\Gamma_{i,j,k} = S_k \cdot Min\left[\delta z_k, 2\frac{<V_j>_{k,z}}{\Omega_k}\right] \qquad [9]$$

To summarize, using Eq.[4] to Eq.[9] it is now possible to compute the encounter velocities for every possible pair of dust sizes in each cell and their corresponding encounter rates.

The evolution of size distributions inside each cell k is then computed using any PIAB algorithm. Several methods exist and are extensively described in the literature (see e.g. Wetherill 1992; Charnoz and Morbidelli, 2004 ; Thebault and Augereau, 2007;  Brauer et al., 2008; Kenyon et al., 2008  etc.). In the current version of LIDT3D, the simple algorithm describes in Brauer et al. (2008) is used. It includes coagulation, fragmentation and craterization, assuming a fixed threshold velocity for fragmentation

**2.5 Step 4: Remapping the system back to the lagrangian representation : the trick of moving tracers (Figure 1.d)**

Let $M'_{j,k}$ the new mass distribution array. It contains, for each cell k, the mass of dust contained in each size bin j at the end of the time-step. Once all $M'_{j,k}$ have been computed the system must be converted back to the lagrangian representation (i.e. using tracers only). The mass of dust in each cell is reattributed to the tracers present in the cell depending on the size bin they belong to. If there is more than one tracer for a given bin size, then the mass of dust corresponding to this bin is equally shared between those tracers. In consequence even though the number of dust grains may increase for a given size bin (because of fragmentation, for example) the number of tracers is kept constant. Only the mass $M_p$ contained in each tracer increases, making these tracers represent a bigger dust cloud (bigger $N_p$).

Difficulties arise when a new population of dust size appears in a cell where no tracer of the corresponding size bin is present. This may happen for many reasons especially at the beginning of a simulation when a steady state has not been reached yet. The coagulation, the fragmentation or the mouvement of grains from one cell to another are good reasons for creating new size populations in a given cell during a single time step. In a protoplanetary disk, for example, this problem occurs when the big grains sediment quickly into the midplane.

To solve this problem, one may be tempted to introduce new tracers in the simulation for the corresponding size bins and cells. However, this would increase the number of tracers, a solution we want to avoid to prevent memory overflow.

The following method is adopted, called a "remapping" (step 4): some tracers are moved instantaneously from one cell k and size bin j to another cell k' and size bin j' where tracers are needed. Although shocking at first sight, this method may be considered simply as an adaptive resolution technic in which the number of tracers is balanced from one cell to another. Indeed, the number of tracers per cell is a quantification of the resolution of the code. Thus, cells in which new particle sizes appear need more resolution to describe their full size distribution. Our method allows us to give them some empty tracers taken from cells that are populated by too many tracers. The latter cells will see their resolution decrease while those in need of tracers will see their resolution increase. Of course this can only be done if the mass is kept constant in each cell and size bin despite the instantaneous reorganization of the tracers. This is achieved in the following way:

First, it is necessary to determine which cells will give and receive tracers and from which to which size bins. To choose the cells k and size bins j that will provide tracers, an optimal number of tracers per cell and size bin ($N_{op,k,j}$) is defined. Many ways of defining $N_{op,k,j}$ may be considered, depending on the way we wish to sample the system. The question that must be answered is: do we need more tracers where there is more mass or do we need about the same number of tracers everywhere in the system in order to better resolve regions that contain less mass? This problem is discussed in the literature of Monte Carlo algorithms, where large systems are represented by a statistical small set of particles, and we refer the reader to this literature (see e.g. Rubino & Tuffin 2009). In the present work, about the same number of tracers per cell and per size bin are needed, so as to conserve a good resolution in the less dense regions. This is necessary in order to compare our results with observations that only detect the disk's upper layers. It is also useful for global disks simulations as particles tend, in their vast majority, to pile up close to the star. With this technic it is possible to redistribute tracers in disk's outer regions despite of their rapid emptying due to gas-drag, and thus, keep a good numerical resolution even in the less populated regions.

So $N_{op,k,j}$ is chosen to be the same for every cell k and size bin j and is thus independent of j and k values. The remapping is done in three steps:

1) The optimal number of tracers per cell k and size bin j ($N_{op,j,k}$) is evaluated as follows :

$$N_{op,k,j} = \frac{N}{\sum_{\text{cells k}} N_{sizeoc}(k)} \qquad [10]$$

$N_{sizeoc}(k)$ the number of size bins in cell k that contain dust grains or in other words, the number of non-zero values in the array $M'_{j,k}$ .

2) The list of tracers that will be moved is selected as follows: For a cell k and a size bin j the quantity $Nt_{k,j} - N_{op,k,j}$ is evaluated (where $Nt_{k,j}$ is the number of tracers in cell k representing dust in size bin j). If it is positive then $Nt_{k,j} - N_{op,k,j}$ tracers may be given to other cells and other size bins. If it is negative then cell k and size bins j needs $N_{op,k,j} - Nt_{k,j}$ tracers. Note that with this method the number of tracers given to other cells is always equal to the number received tracers in order to preserve constant the total number of tracers in the system. The list of tracers that will be moved from cells k and size bin j to their target cell k' and size bin j' is then determined randomly.

3) All shifted tracers are given a new position and velocity vector (see below) inside their target cell k'.

4) Once the required tracers have been moved, all of the grid's cells contain the optimal number of tracers to hold the new mass size distribution (the M'$_{j,k}$ computed after the collisional evolution). In each cell the dust of each range of size is spread equally in the tracers of the corresponding size bin.

5) For all cells k, the dust-mass distribution is distributed equally for every size-bin j in all the tracers they contain

The last thing to describe is how the new positions and velocities of tracers are computed when moved to a new cell and size bin. Several methods were tested and it is very possible that the best method depends on the considered problem. The following simple rule produced good results for dust in protoplanetary disks.

1) Draw a position in cell k at random with a uniform distribution: let **r$_s$** the position vector

2) Build the velocity vector **v$_s$** as follows:

$$\vec{V}_s = \vec{V}_g(\vec{r}_s) + <\vec{V}_j>_k - \vec{V}_{k,g} \qquad [11]$$

Where **V$_g$**(r$_s$) is the gas velocity at **r$_s$**. For a gas-free disk **V$_g$** may be self-consistently replaced by the local circular velocity. Eq.[11] states the starting velocity vector is the gas velocity plus the average difference between the gas velocity and the average velocity of particles of size j. In the case <V$_j$>$_k$ is not determined (because there was no dust in the size bin j before) then <V$_{j''}$>$_k$ is used instead, where j'' is the closest bin to j for which <V$_{j''}$>$_k$ is known. This procedure is very adapted to a protoplanetary disks because gas-drag makes all particles move at the same velocity thus making any velocity dispersion between particles of the same size close to 0. An optional method may also be used so as to keep track of the velocity dispersion:

$$\vec{V}_s = \vec{V}_g(\vec{r}_s) + <\vec{V}_j>_k - \vec{V}_{k,g} + \begin{pmatrix} W_r \\ W_\theta \\ W_z \end{pmatrix} \cdot \begin{pmatrix} \sigma_{r,j,k} \\ \sigma_{\theta,j,k} \\ \sigma_{z,j,k} \end{pmatrix} \qquad [12]$$

Where W$_r$, W$_\theta$, W$_z$ are three independent random numbers with a normal distribution and $\sigma_{r,j,k}$, $\sigma_{\theta,j,k}$, $\sigma_{z,j,k}$ are the 1 sigma standard deviation of the velocity vectors of particles in size bin j in cell k, in the three directions radial (r), azimuthal (θ) and vertical (z). Eq. [11] and Eq.[12] gave similar results for a protoplanetary disk.

Contrary to the case of protoplanetary disks that are rich in gas, our algorithm to define the starting velocity of a moved tracer is not satisfying in debris disks. In protoplanetary disks the dust's dynamics is mostly controlled by gas drag thus particles quickly loose the memory of their starting velocity vector. In debris disks though, particles tend to keep the memory of their initial velocity vectors because of the absence of a dissipative medium. The initial velocity vector must be defined with special care. This limitation is under investigation and our findings will be presented in a coming paper.

For the moment our procedure does not strictly conserve energy (E) nor angular momentum (J) and this is clearly a weakness. However, in the context a protoplanetary disk, the gas acts as an infinite source of both energy and angular moment as it gives/takes E and J from/to the dust grains and does not back-react to the presence of the grains. So it is not possible for the moment of simulate things like streaming instability for example. The terminal velocity of the grains is imposed by the gas-dynamics and is independent of the initial E and J of particles (after several coupling timescales). Since the terminal velocity is naturally correctly computed in our model (due to the direct integration of tracers' motion) this approximation does not appear to be too wrong. A possible improvement would be the following: as mass is shifted from one tracer to another, E and J may be moved fomr one tracer to another in the same fashion as the mass. This improvement will be implemented in a close future.

The trick of moving tracers implies also a loss of the lagrangian property of the code. Since tracers are sometime moved from one place to another it is impossible (or at least, very difficult) to reconstruct the trajectory of an individual tracer. However this loss of information is not vital for the global evolution of the system. A way to avoid this is to define some tracers that may be shifted and some others that could not. Of course the former must be much more numerous than the latter.

The redistribution of tracers has been partially parallelised (in open-mp). It is done in the following way

1. All processor compute independently for each cell the number of tracers that may be exchanged (with other cells) and establish the list of those tracer (one list per cell).
2. One processor gathers all the lists from every cell and create a unique list. Then it reads this list and attributes the exchangeable tracers to their new cell and de-attributes them from their starting cell.
3. All processors examine independently all cells and recomputed the mass content and orbit of all tracers they contain.

The phase of de-attribution and re-attribution from cell to cell could not be parallelised because it implies to modify elements in an array in an unpredictable way… so that parallelising this action may lead to collisions between processors modifying, from time to time, the same elements in an array . For the moment this bottleneck is not solved.

**3. Application example within a protoplanetary disk**

A local simulation at 1AU is performed and the evolution of the dust size-distribution is computed self-consistently at any distance from the midplane The effect of gas-drag, turbulence as well as coagulation and fragmentation are coupled. The dynamical effect of collisions (dissipative rebounds) is neglected for the moment and will be included in a future work. However, since the dust dynamics is strongly controlled by the gas, dust-grains trajectories reach rapidly a terminal velocity (so there is a rapid loss of memory of the dust's velocities, just after an impact). So for bodies with short gas-coupling timescales (i.e. with stokes number < 1) the absence of dissipative rebounds should not be a problem. However for big bodies (stokes number >1) or when the dust/gas ratio > 1 this approximation may fail. It will be treated in a forthcoming paper.

**3.1 Local simulation: the code step-by-step**

As an illustrative example a local simulation is run in a ring extending from 1 to 1.1 AU with radial periodic boundary conditions. The initial dust size distribution is typical of the interstellar Medium. The four steps described in section 2 are illustrated below :

- Step 1: the tracers' motion is integrated from time t to time t+dt (Figure 2). A classic implicit ODE solver is used to integrate the Eq [1].

- Step 2: a grid is defined on top of the system (Figure 3). The cells from left to right have not exactly the same vertical dimension because this dimension scales with the local pressure scale height, so that there is always about the same number of cell per pressure scale-height. In each of the grid's cell, the velocity and mass distributions are extracted summing the dust content of each tracer. As an example, in Figure 4 are presented the mass and velocity distributions of dust grains in a cell close to the midplane: cell #3 (see Figure 3). This cell contains dust grains up to $10^{-4}$-$10^{-3}$ m radius that have sedimented in the midplane (Figure 4.a). The velocity distribution is computed from the tracers' velocity vectors associated to the different size bins (Figure 4.b). It clearly shows that these particles migrate radially inward (negative radial velocity) and sediment with a vertical velocity that is an increasing function of their size. By comparison, in cell #17 (Figure 5.a) that is at a much higher altitude (see Figure 3), the mass distribution is poor in particles larger than $10^{-4}$ m (because of sedimentation). The velocity distribution of dust inside cell #17 (Figure 5.b) shows a super-keplerian azimuthal velocity. This known feature of vertically extended dust-disk above ~1.5 pressure scale height is consistent with Takeuchi & Lin (2002). The effect of these particles that drift outward has never been investigated before. In this respect our code opens new perspectives as it allows the study of particles in any arbitrary dynamic state. Note also that the vertical velocity of particles in cell #17 is larger than in cell #3. This will make the collisional activity in the disk's upper layers more intense than close to the midplane.

- Step 3: the collisional evolution is computed in each cell using a standard PIAB collisional code. This code uses the collision rates computed with Eq.[8] and encounter velocities computed with Eq.[7].

- Step 4: In each cell the tracers are filled with the dust corresponding to their size bins. If there is no tracer in a given size bin in some cells, tracers are taken from cells that have too much tracers and given to those that lack tracers. In Figure 6 arrows show the starting and ending positions of all tracers that were exchanged during the time-step. Note that cells close to the midplane tend to "give" tracers to cells high above the midplane. This is due to sedimentation that makes the dust move toward the midplane. As a result the upper layers of the disk lack tracers. The remapping technique balances this effect and redistributes the tracers equally in every cell.

**3.2 A Test: comparison with previous results**

It is not straightforward to compare our result with the previous ones because the method used here, and a large part of the physics considered in LIDT3D, are intrinsically different from the previous studies. One of the principal advantages of our method is that it frees us from several usual approximations. In previous analyses we were obligated, for example, to use simple prescriptions for the dynamics, or instantaneous diffusion for the grains, or else treat the dynamics close to the midplane only. This advantage of the method is also one of its drawbacks since in case of a discrepancy between results it may be difficult to disentangle an error in our code from a bad approximation. To our knowledge, the closest approach to our code is the one presented in the work of Brauer et al.(2008) and Birnstiel et al. (2009). They use an analytical prescription of the dust dynamics valid in the disk's midplane where they assume an instantaneous gaussian vertical distribution. Encounter velocities between particles take into account the particle's sedimentation, radial drift, turbulent forcing and thermal motion. Several cases are considered in Brauer et al. (2008) in which different physical processes are considered (radial drift, vertical sedimentation, coagulation, fragmentation). To test our code a case in which all processes are included is considered and corresponds to the case presented in section 3.3 and Figure 13 of Brauer et al. (2008). A transition disk is considered (with a gas surface density at 1AU of about 200 kg/m$^2$, rather than 30x10$^3$ kg/m$^2$ for a MMSN) surrounding a 0.5 solar mass star, and the value of the $\alpha$ parameter of turbulence is 10$^{-4}$ (quite low) . The critical fragmentation velocity is set to 10 m/s (quite high). Simulations are performed for R=1AU and 10AU. Turbulence is included in the form of a stochastic forcing on particles position (see e.g. Charnoz et al., 2011; Ciesla et al., 2010). Relative velocities due to turbulence are computed through Eq.[7] using the formalism of Ormel and Cuzzi (2007). Coagulation and fragmentation laws considered here are very simple (adapted from Brauer et al., 2008). Coagulation occurs when relative velocities $V_{rel}$ are lower than $V_{frag}$ and $V_{frag}$=10m/s is assumed. If $V_{rel}$>$V_{frag}$ and if the mass ratio (smallest mass divided by the largest one) of the two colliding particles is larger than 0.1 then catastrophic fragmentation occurs. Fragments are distributed with a size distribution following dN/dr$\propto$r$^{-3.5}$ (starting from particles with radius 0.1μm and conserving mass). If the mass ratio is smaller than 0.1 then cratering occurs with the largest fragment having the same mass as the impacting body.  Of course these laws are simple and may be somewhat unrealistic however they capture the main ingredients of a collisional evolution model. They will be replaced in the future with a more detailed fragmentation as in Thebault and Augereau (2007) or in Birnstiel et al. (2010).

The resulting mass distributions (averaged over the full disk thickness) after 10$^4$ years of evolution is shown in Figure 8 and should be compared with Figure 13 of Brauer et al. (2008). First note that the mass distributions are remarkably similar: flat with a sharp cut-off. This feature is due to the collision velocity that increases (because of the gas drag) with the dust size.  When the drift velocity is comparable to the fragmentation threshold velocity, coagulation is no longer possible.  The cut-off size at 1 and 10 AU is about 2 mm and 0.8 mm respectively, in very good agreement with Brauer et al. (2008).  The fact that in the most complete simulation, including all the effects, a good agreement, is found is, in some sense, a cross validation of both works. On the other hand since far less approximations are performed in our case, we are comforted in the validity of our approach.

### 3.3 Example of application: particle growth with and without a dead zone

As an example of the useful potential applications of LIDT3D, we consider the problem of dust growth in a protoplanetary disk. The nfluence of altitude and vertical stratification on dust growth is of particular interest. One of the outstanding problems of planetary formation is to understand how dust grains may grow into kilometic bodies. In Brauer et al. (2008) it is shown that particles may grow to a few millimeters in size, assuming a disk quite favorable to accretion: a low $\alpha$ value (10$^{-4}$) and quite a high velocity threshold for fragmentation (10 m/s). However, the value of $\alpha$ depends both on the disk's mean magnetic field and on its ionization state. An average value of  5x10$^{-3}$ is found in Fromang and Nelson (2009), however this value may vary from 10$^{-4}$ to 10$^{-1}$ depending on the conditions.  and that the fragmentation velocity is rather in the range cm/s to m/s (Blum and Wurm 2008) for particles larger than a micron. Recent theoretical developments have shown that the region close to the disk's midplane may be neutral, and as such not subject to magnetically driven instability (MRI). This region, called the "dead zone" may be almost laminar with a low value of $\alpha$. In this region, coagulation is expected to be more efficient due to lower encounter velocities between particles.

Such a disk may be challenging to model because of vertically stratified dynamics. It was out of reach of all previous codes because of a Z dependent dynamic induced by a vertically varying $\alpha$. As an illustration it is demonstrated that LIDT3D is able to tackle this problem. We wish to address the following questions:

- How may the presence of a dead zone modify the grain size distribution and the global growth of the dust?

- How are dust grains distributed vertically, with and without a dead zone?

A turbulent gaseous protoplanetary disk is considered, with the properties of a minimum-mass solar nebula: at 1 AU the surface density is $\sigma_0=30\times10^4$ kg/m$^2$ (about two times the MMSN, because it is recognized not to be massive enough to make today's planets, however note that it is quite arbitrary), temperature $T_0=280$ K, gas scale height $H_0=0.03$ AU, the initial dust/gas mass ratio is 1% (see Takeuchi and Lin, 2002). 10000 tracers are introduced, representing initial dust-particles with radii ranging from 0.1µm to 10µm and a differential size distribution $dN/dr \propto r^{-3.5}$. The spatial grid is paved with 60 cells (2x30). 2 turbulence prescriptions are considered: for the disk without a dead zone, $\alpha=5\times10^{-3}$ throughout the disk's thickness, while for the disk containing a dead zone, $\alpha=5\times10^{-6}$ between Z=-1.5H to Z=1.5H (the dead-zone) and $\alpha=5\times10^{-3}$ in the active layers above. The fragmentation threshold velocity is set to 1 m/s, in agreement with Blum and Wurm (2008).

The resulting size distributions are compared in the top of Figure 9 after 2000 years of evolution, when a steady state is reached. In both cases the mass of the system is mainly contained in the largest bodies and the distributions present similarities with those of Brauer et al. (2008). In the presence of a dead zone, the dust grows two orders of magnitude larger. Indeed, in the case without a dead zone the maximum size reached is about 0.8mm whereas in the case with a dead zone it is 8 cm. This is due to the lower encounter velocities inside the dead zone. Furthermore, the objects that reach the dead zone are then trapped efficiently in response to the low viscosity (which corresponds to a low turbulent diffusion coefficient). The vertical distributions of dust are presented in Figure 10. Without a dead zone, they are mostly Gaussian features, apart from the dominant particle size that seems somewhat depleted in the midplane, because of very efficient fragmentation. In the presence of a dead zone particles larger than 1mm are very abundant and their density may even be greater than that of micronic grains close to the midplane. It is also instructive to compare the size distributions above the dead zone (in the active layers region) and inside the dead zone. This is presented in Figure 9.b. The blue line shows the size distribution inside the dead zone (<1.5 pressure scale height). It is very similar to the vertically integrated mass distribution of the full disk, except for dust smaller than 0.2 mm that is clearly depleted. Conversely, dust in the active layers (red line, above 1.5 scale height) consist mostly of particles up to 0.1 mm but is largely depleted in big bodies because of sedimentation and diffusion. Figure 9 strongly suggests that the presence of millimeter to centimeter sized particles may be a signature of a dead zone. However the main differences appear for particles larger than 0.1 mm , and thus, radio-observation may decipher the presence of a dead zone. In this respect the forthcoming ALMA facility will provide very interesting opportunities to characterize the occurrence of dead zones in protoplanetary disks. The vertical structure of these disks will be studied in details in a forthcoming paper, as the present figures only illustrates the capabilities of LIDT3D.

## 4. Conclusion

We have presented a hybrid algorithm to couple self-consistently the collisional and dynamical evolution of dust in a circumstellar disks with a special emphasis on the case of a protoplanetary disks. The key features of the code are the following:

- The use of "super-particles" (the tracers) to represent a cloud of many dust grains with a single size rather than a single body.

- The use of a "collisional grid" on top of the system inside the cells of which the evolution of size distributions are computed

- A "remapping" algorithm to reorganize tracers at each time-step.

The last aspect is the most challenging to put into practice: at each time step the system is mapped back-and-forth from a lagrangian representation using tracers (to compute the dynamical evolution and derive encounter velocity) and a eulerian representation using the collisional grid (to compute the collisional evolution using any Particle in a Box approach). Tracers, which are fundamental elements of this approach, can be moved at need at each time-step in order to optimize the mapping of the current spatial distribution of dust. A nice aspect of this approach is that the number of tracers is kept constant, making the computation of the system manageable despite the increasing number of dust grains. This method has already been successful in the case of a protoplanetary disk and its application to a debris disk is still under development. In the case of a debris disk it seems that our algorithm to define the new velocity vectors of tracers after the remapping phase (step 4, see section 2.5) is not very satisfactory and work is still needed to adapt LIDT3D to the case of debris disks, however there is no intrinsic limitation.

The hybrid technique presented here is quite flexible and may be used to take additional perturbers into account (like a planet or a companion star) as there is no *a priori* assumption on the dynamics, contrary to the majority of works up to now.

An application example relevant to the case of a protoplanetary disk was presented: the growth of dust in a ring of a gaseous protoplanetary disk at 1 AU with and without a dead zone. For the first time the vertical distribution of dust was self-consistently computed. It is found that that with a dead zone the maximum dust grains size may be up to 100 times larger than without. Major differences between the two cases appear for grains > 0.1 mm, that may be detectable in radio emission. The possibility to grow particles up to centimetric sizes inside a dead zone is particularly appealing because it is about the size required for grains to efficiently decouple from the gas and accumulate inside large scale vortexes where they may grow up to a hundred kilometer in size as shown in several recent works (Johansen et al., 2007; Cuzzi et al., 2008; Baie and Stone 2011). So it may be possible that the growth of dust into planetesimals or planetary embryos may be a two steps process:

- Growth from sub-micronic to centimetric sizes via coagulation in dead zones

- Growth from centimeter to > 10 km sizes either in large-scale pressure bumps (Johansen et al., 2007) or between small-scale eddies near the Kolmogorov scale of turbulence (Cuzzi et al., 2008), or in particle pile-ups caused by streaming instabilities (Bai & Stone, 2010).

This new technique opens new potential applications, such as the formation and tracking of CAi during the evolution of the protoplanetary disks, the determination of observational tests to infer the presence, or not, of a dead-zone or of planetary sized bodies in a distant protoplanetary disks, etc.

Future improvement should include in priority the taking into account of the dynamical effect of collisions, in the form of dissipative collisions. In principle this could be included in the phase of orbits-redefinition, in step 4 (orbits may be modified to simulate the net energy loss). In addition, strict conservation of energy is not realized for the moment. This should be also corrected in step 4 where orbits can be slightly modified to ensure strict conservation of energy when moving tracers from one cell to another.

**ACKNOWLEDGEMENTS :**

S.C. wishes to thanks Philippe Thébault and Neal Turner for enlightening discussions and an anonymous referee whose comments have largely contributed to increase the quality of the paper. This work was supported by a grant from the "Campus Spatial" program of Université Paris Diderot, by the CEA/IrFU/Sap, as well as by the IUF (Institut Universitaire de France). E.T. is supported by a "Région Ile de France" grant. We wish to thank Jacques Labbé for his linguistic help.

# FIGURES

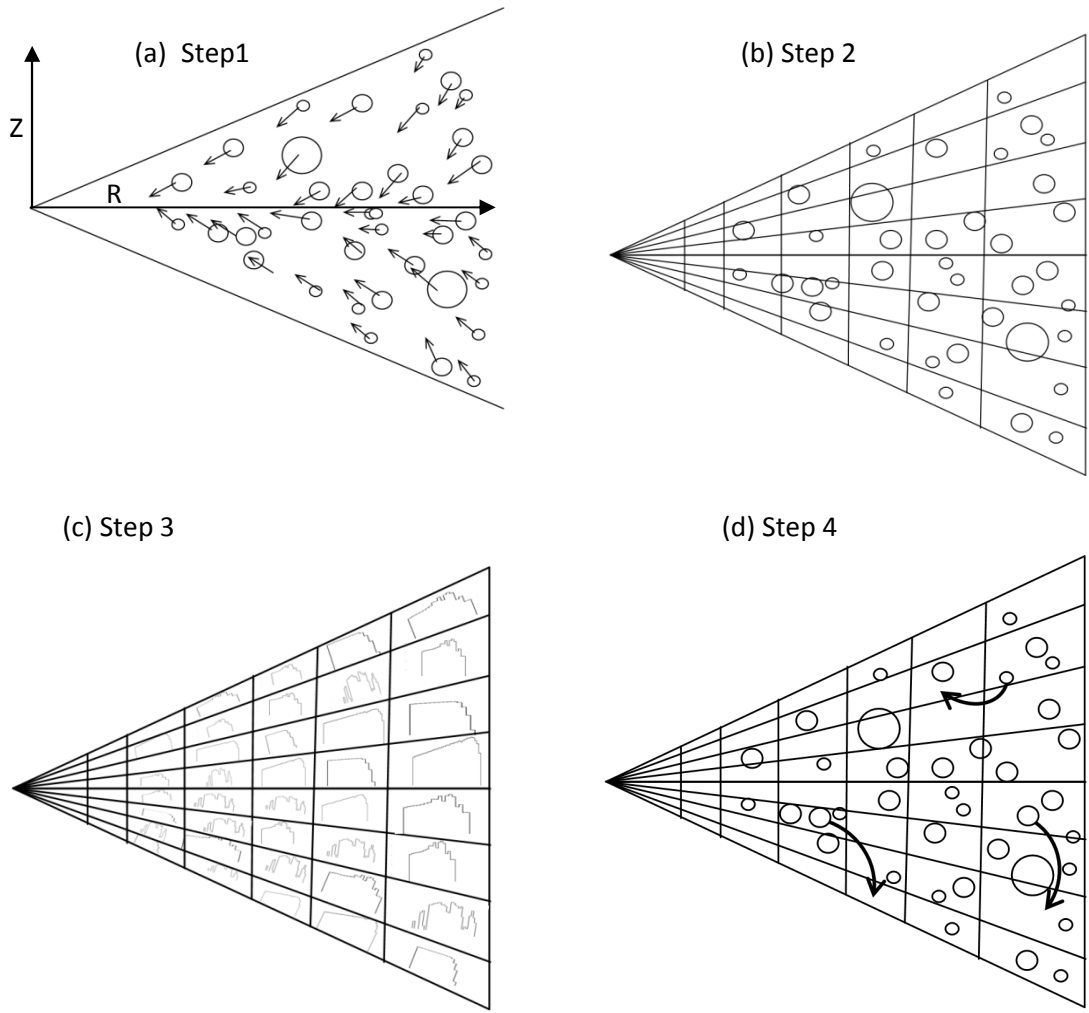

Figure 1 : A typical time-step in LIDT3D. (a) Step 1: integration of the tracer's motion in three dimensions (b) Step 2: remapping the tracers into a grid and computation of local encounter velocities (c) Step 3: evolution of the size distributions in each grid cell using a Particle in a Box algorithm (d) Step 4: remapping the size distributions contained in each cell into tracers and spatial reorganization of tracers so that there is about the same number of tracers per size bin per cell. If tracers of a given size are lacking in one cell, we move a tracer from another cell (which has too many tracers) and give its dust content to the neighboring tracers of the same cell and size bin prior to the shift to the new cell.

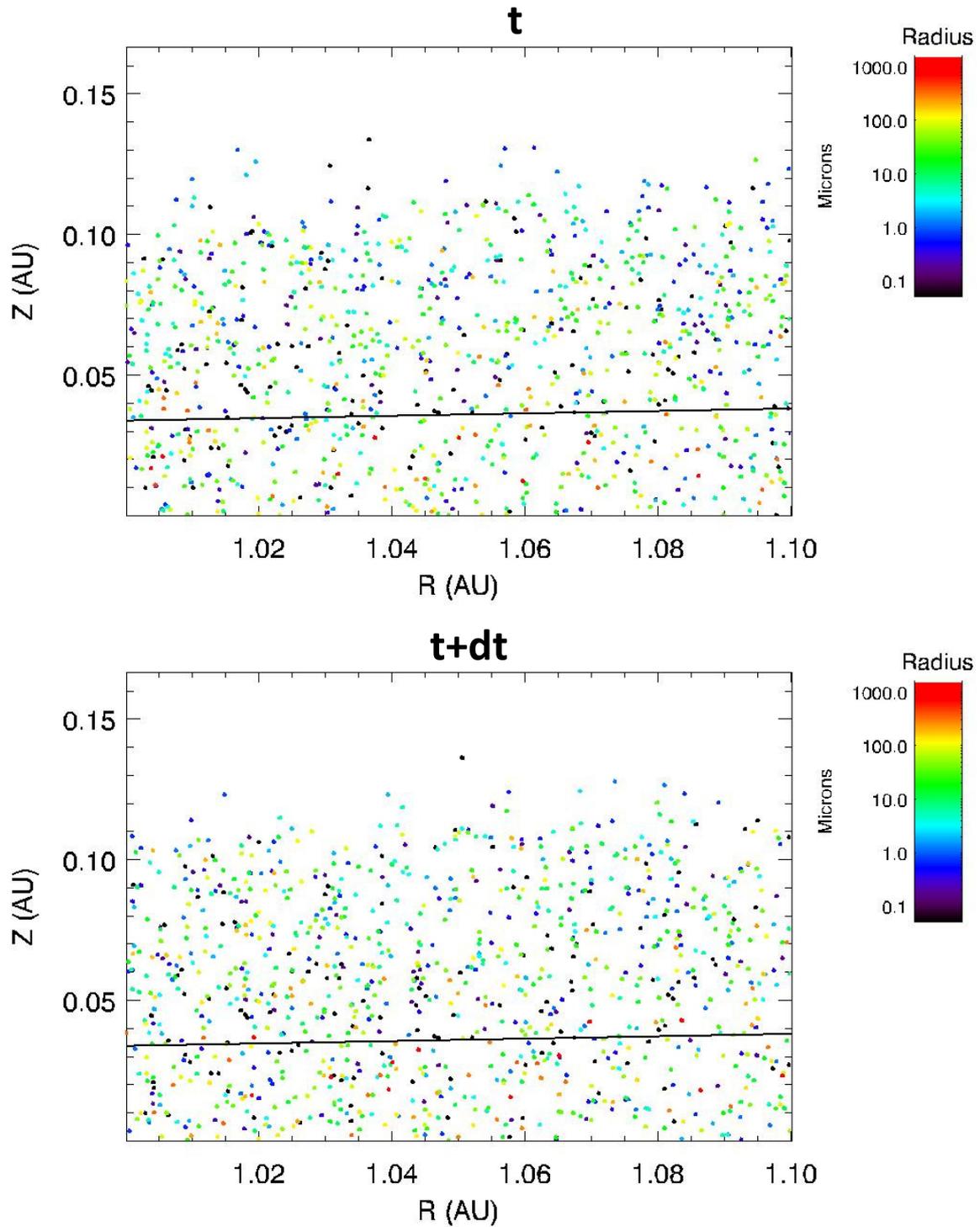

**Figure 2 : Step 1** : tracer motion is integrated from t to t+dt (dt=1 year here). We consider a local simulation in a ring extending radially from 1 to 1.1 AU with radial periodic boundary conditions. Gas drag and turbulence are included in the simulation. The solid line shows the pressure scale height (H). Each point represents a tracer (1000 tracers in total here), and the color designates the average size of the dust grains contained in each tracer. Top: system at time t; bottom: system at time t+dt.

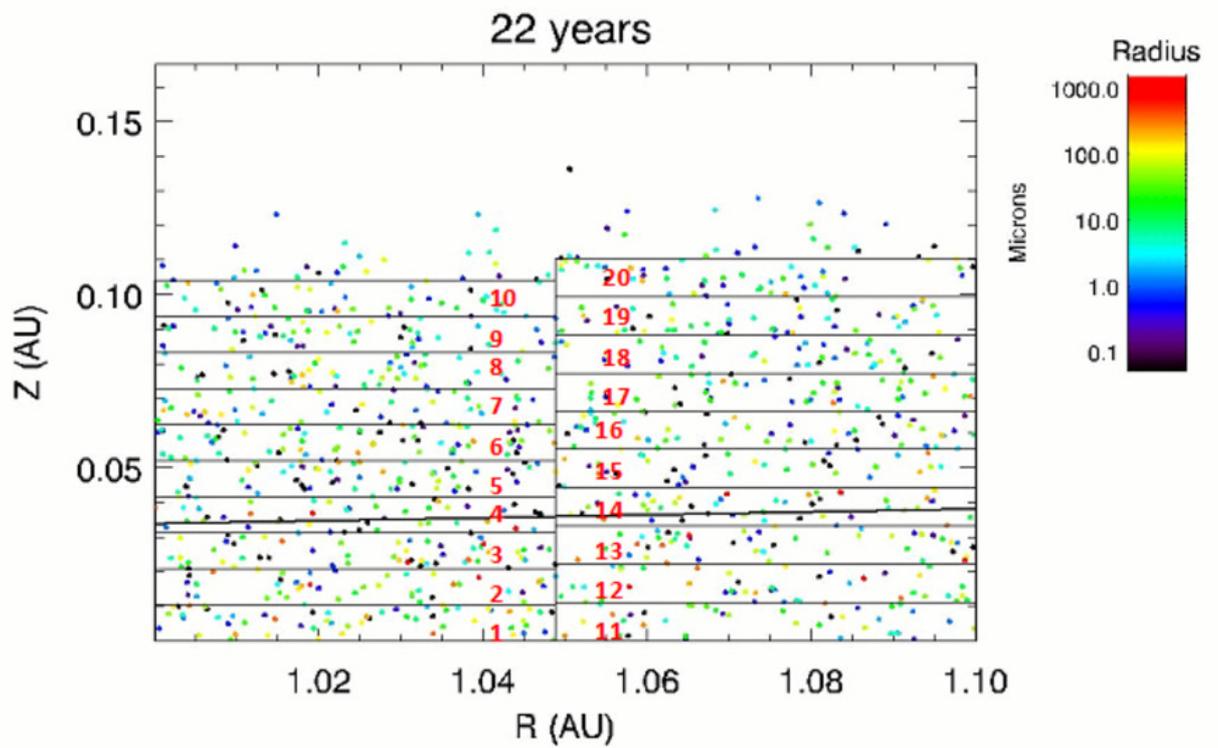

**Figure 3 : Step 2** : a 2D grid is put on top of the system. In each cell, the average encounter velocities and the local size distribution are computed.. Here there are 10x2 cells in the grid. The bold solid line represents the scale height of the gas disk. The numbers in red designate the index number k of each cell.

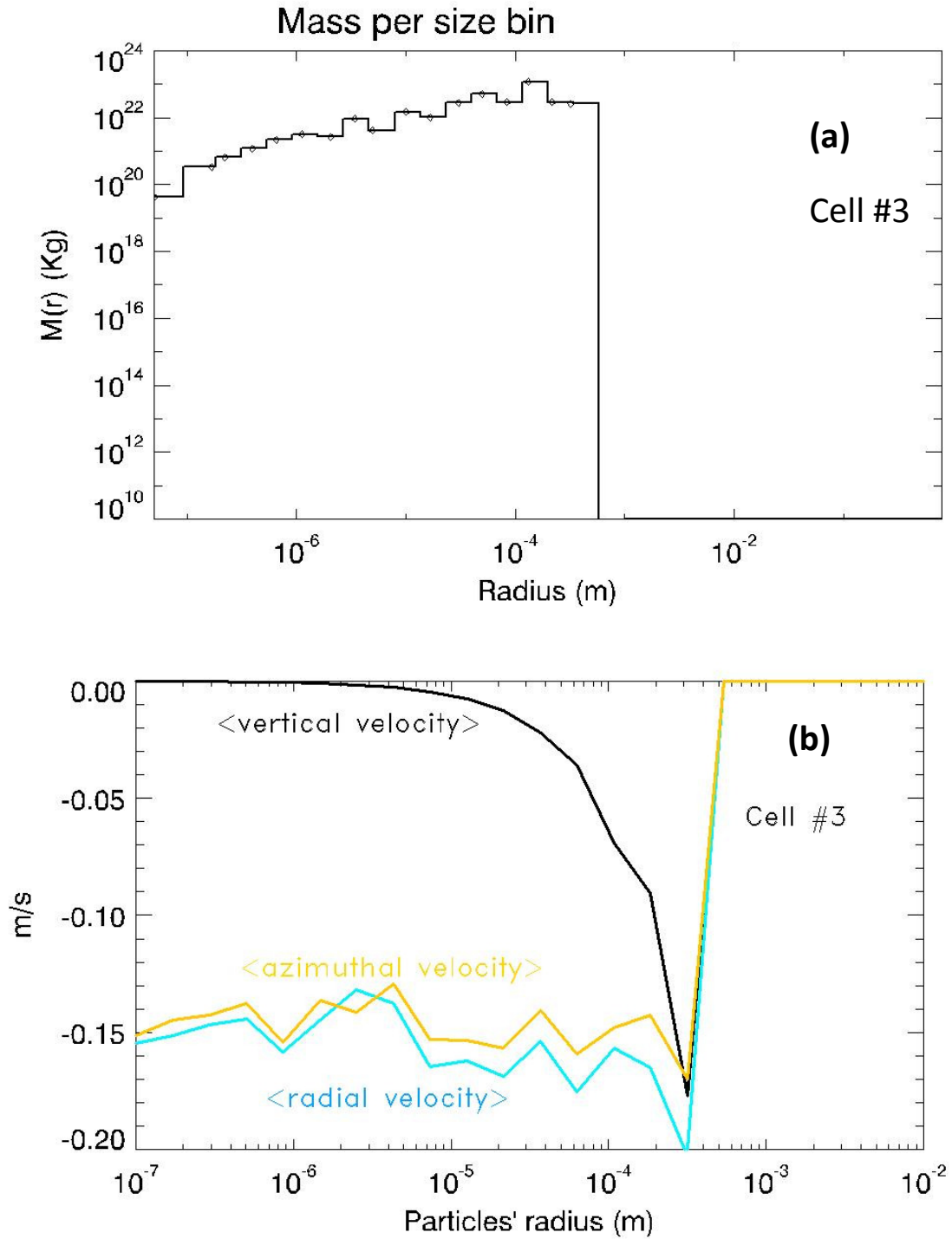

**Figure 4 :** Illustration of **step 2**: Computation of the velocity distributions for each particle size in every cell. Here we show the content of Cell #3 which is close to the midplane (see Figure 3). (a) shows the mass distribution of dust in cell #3. (b) shows the average velocity vector (in the inertial frame) as a function of particles' radius. Blue line: radial velocity, yellow line: azimuthal velocity, black line: vertical velocity.

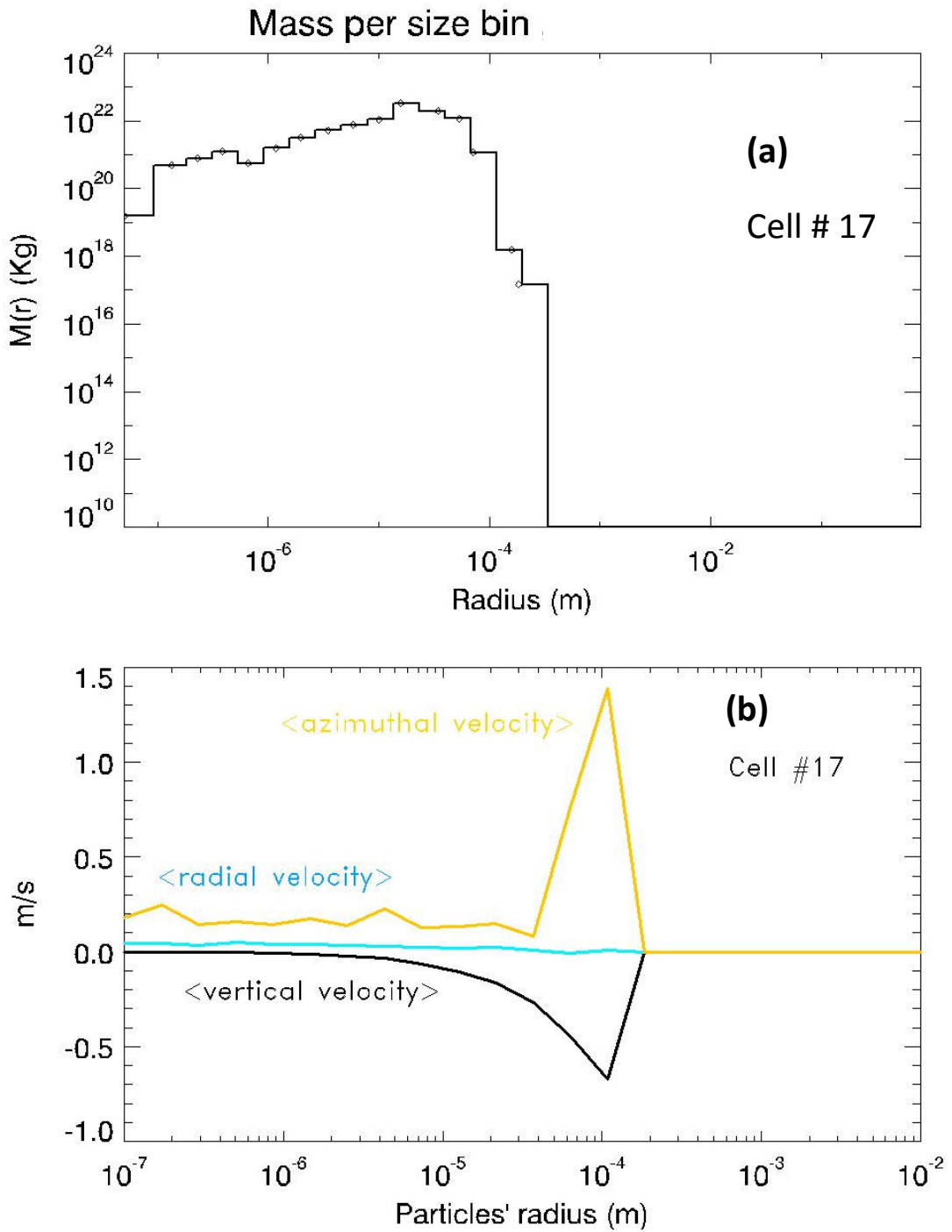

**Figure 5 :** Same as Figure 4 for cell #17 which is far from the midplane (see Figure 3).

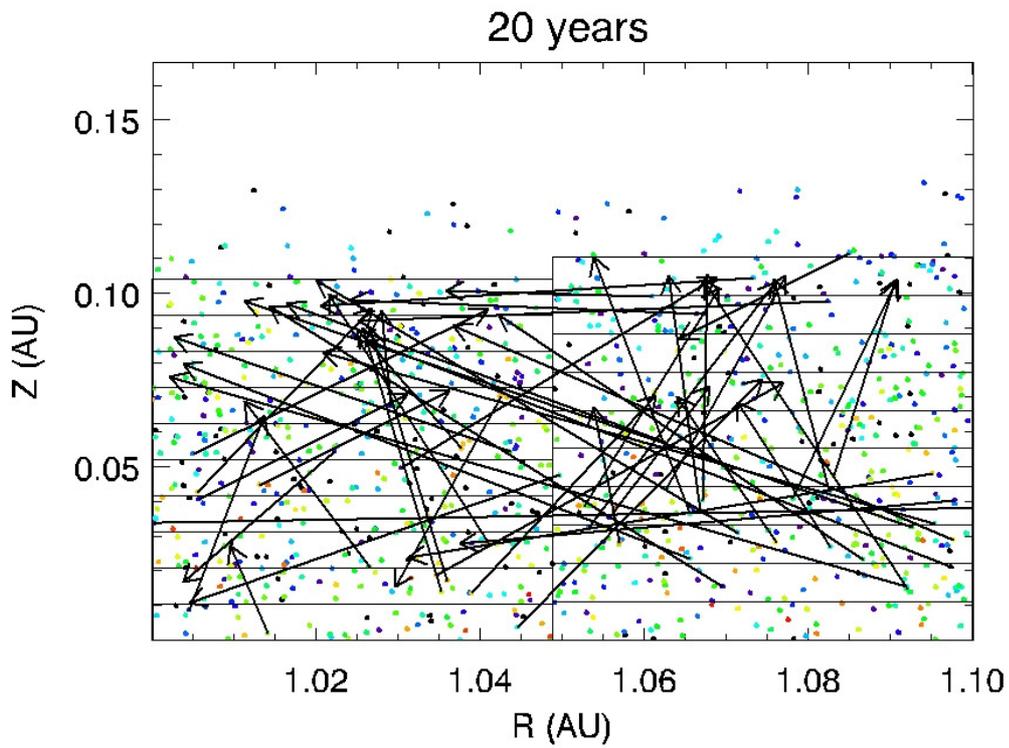

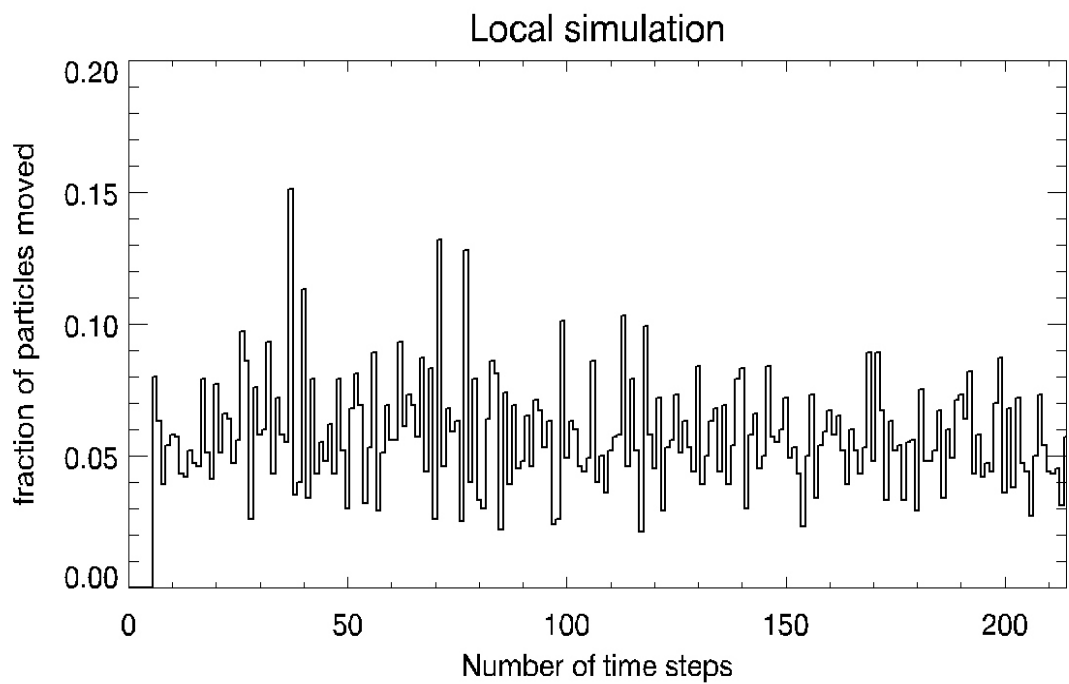

**Figure 6 :** STEP 4 : some tracers are shifted from one cell to another so that every cell has enough tracers of all sizes to represent the dust size distribution they contain. Top: example of shifted tracers: here the arrows' bases designate the tracers that were moved in other cells (at time t) while the arrows' heads designate the location at which they were moved just before beginning time-step t+dt. Particles out of the grid do not participate in the collisional evolution. Bottom: proportion of tracers shifted at each time step.

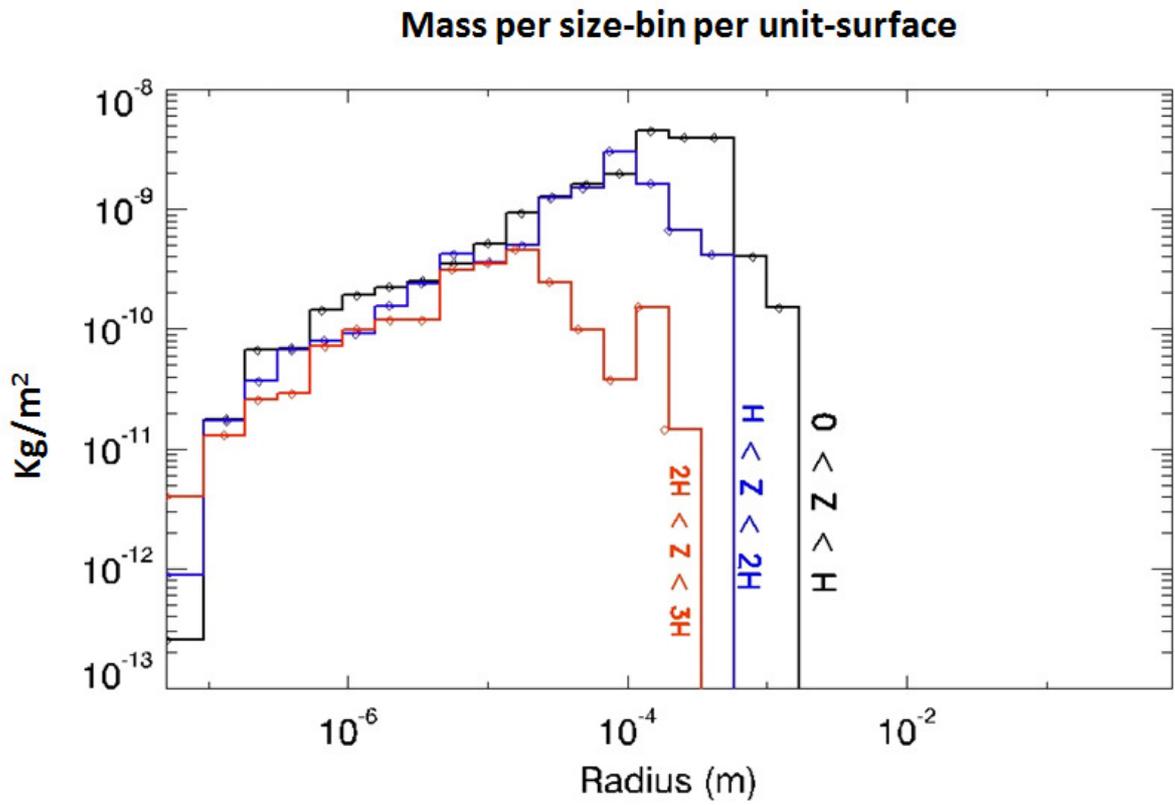

**Figure 7 :** Mass distribution in the disk at different scale heights. Diamonds represent the mean dust size in each size bin. Black line: mass distribution of dust contained between Z=0 and Z=H, blue: mass distribution of dust contained between Z=H and Z=2H , red : for 2H<Z<3H. The effect of sedimentation is clearly visible and small dust grains are more abundant in the disk's upper layer while there is a deficit of big particles.

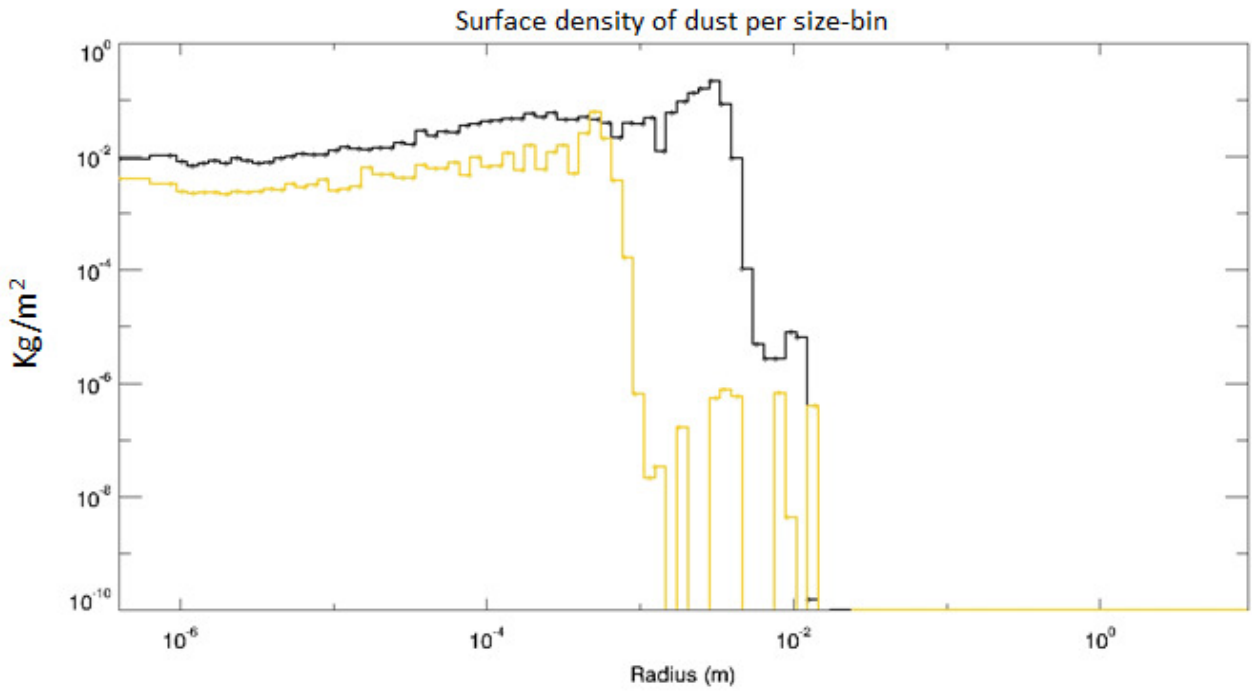

**Figure 8** : Particle size distribution at 1 AU (black line) and 10 AU (orange line) after $10^4$ years of evolution obtained with LIDT3D in a transition disk as in Brauer et al. (2008) including coagulation, fragmentation and turbulence. These results closely match those in Figure 13 of Brauer et al. (2008).

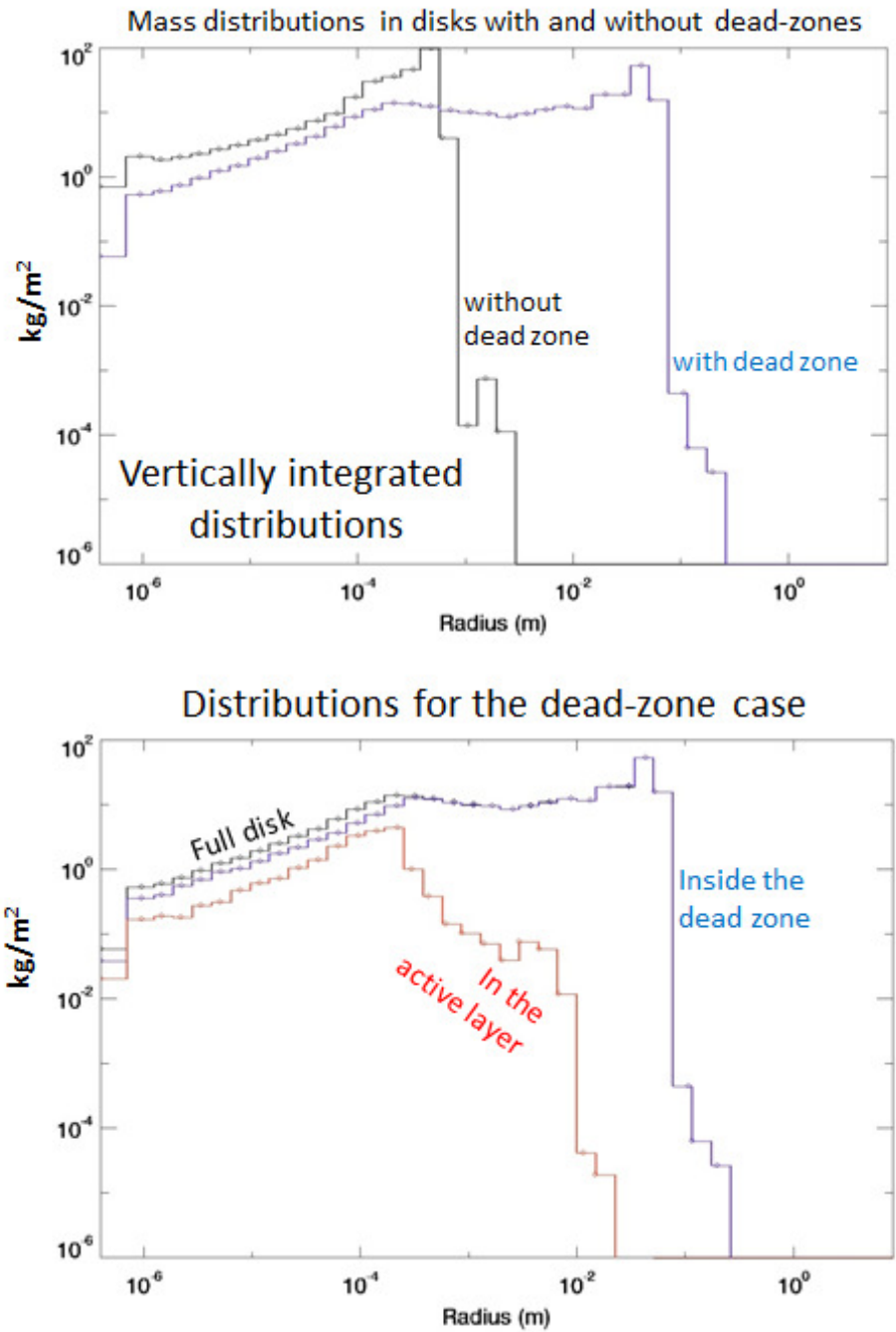

**Figure 9 :** Vertically averaged dust mass distribution (surface density in each dust size bin) in a disk (MMSN) at 1AU in disks with and without dead zones. In the dead zone, $\alpha=5\times10^{-6}$ and in the active layer (top of the dead-zone) $\alpha=5\times10^{-3}$. Top: black line: no dead-zone, blue line: with dead zone. Bottom image: black line: mass distribution vertically averaged, blue line: distribution in the dead zone only ($\pm1.5$ H), red line : distribution in the active layer (Z>1.5H).

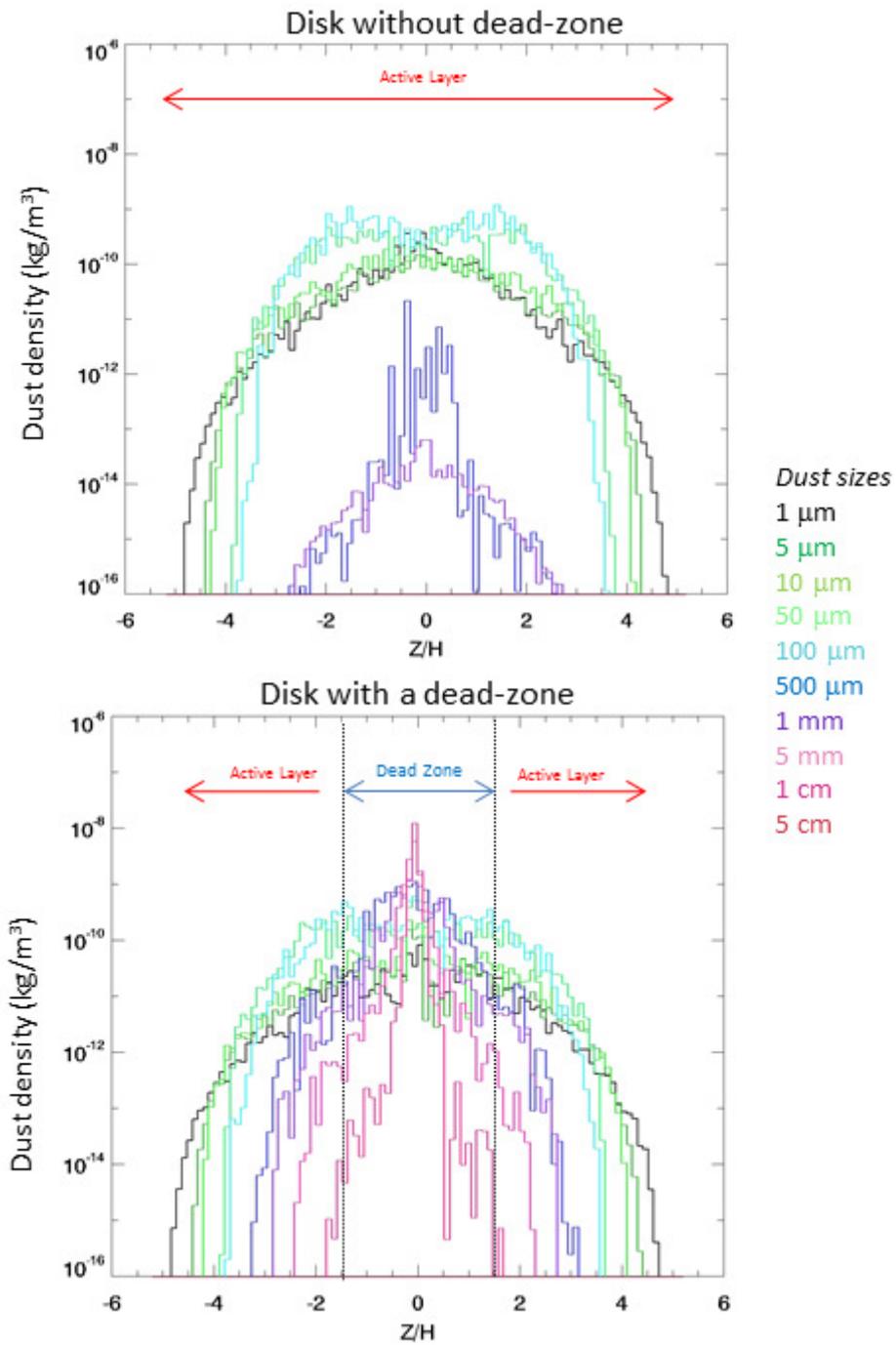

**Figure 10 :** Vertical distribution of dust (in kg/m$^3$) in the case of a disk without a dead zone (top) and of a disk with a dead zone between ±1.5H (bottom). The colors designate the dust size.